\documentclass[]{article}
\usepackage{fullpage}
\usepackage{graphicx}
\usepackage{subfig}
\usepackage{latexsym}

\title{Particle Identifications from Symmetries of Braided Ribbon Network Invariants}
\author{Sundance Bilson-Thompson\thanks{Email address:
sbilson-thompson@perimeterinstitute.ca}\\
Perimeter Institute for Theoretical Physics,\\
31 Caroline St. N., Waterloo, Ontario N2L 2Y5, Canada,
 \and Jonathan Hackett\thanks{Email address:
jhackett@perimeterinstitute.ca}\\
Perimeter Institute for Theoretical Physics,\\
31 Caroline St. N., Waterloo, Ontario N2L 2Y5, Canada, and \\
Department of Physics, University of Waterloo,\\
Waterloo, Ontario N2J 2W9, Canada\\
 \and Lou Kauffman\thanks{Email address:
kauffman@uic.edu}\\
University of Illinois at Chicago\\
750 S Halsted St\\
Chicago, IL 60607, United States\\
 \and and Lee Smolin\thanks{Email address:
lsmolin@perimeterinstitute.ca}\\Perimeter Institute for Theoretical Physics,\\
31 Caroline St. N., Waterloo, Ontario N2L 2Y5, Canada, and \\
Department of Physics, University of Waterloo,\\
Waterloo, Ontario N2J 2W9, Canada\\}
\begin{document}

\maketitle

\begin{abstract}

We develop the idea that the particles of the standard model may arise from excitations of quantum geometry.  A previously proposed topological model of preons is developed so that it incorporates an unbounded number of generations.   A condition is also found on quantum gravity dynamics necessary for the interactions of the standard model to emerge.

\end{abstract}

\section{Introduction}

Preon models, in which quarks and leptons are composites of more elementary 
preons, were developed in the 
1970s\cite{Pati:1974yy,Terazawa:1980hh,Harari:1979gi,Shupe:1979fv,Harari:1980ez}  
and then largely abandoned because they could not be made to work within a 
conventional framework based on Yang-Mills fields coupled to chiral fermions.  
Even in the most promising class of preon models there appeared to be a 
need for either an unusual quantum number or unusual statistics that kept track of 
something like ``position" within the bound state, and there were strong 
constraints on chiral fermions resulting from bound states of preons.

A few years ago the first problem was solved by coding a preon model in terms of braided ribbon networks \cite{Bilson-Thompson:2005bz}.  The coding using a braid naturally incorporates the unusual quantum number required by the preon models.  This made possible a proposal for a solution of the second problem, which is that the quarks and leptons are topological excitations of quantum geometry.    In \cite{Bilson-Thompson:2006yc} it was shown that in a simple model of quantum gravity inspired by loop quantum gravity, there are topologically conserved excitations, the simplest of which correspond to the topological coding of the preon model given in \cite{Bilson-Thompson:2005bz}.  This realized the proposal\cite{fot-dav} that elementary particles might arise from noiseless subsystems\cite{nfs} in background independent approaches to quantum gravity.

The results of \cite{Bilson-Thompson:2006yc} suffered, however, from two limitations, which motivated the work reported in this paper.  First, the correspondence to the standard model was restricted to the first generation.  Here a correspondence with the standard model is  continued to an unbounded series of generations.  This is the main result of this paper and also constitutes a prediction of the unification scheme studied here: there are more than three generations of standard model fermions.

The second issue is that, as demonstrated in \cite{Hackett:2007dx} the excitations we would like to identify with the standard model fermions are conserved too strongly: each constitutes a separate noiseless subsystem, hence while there will be scattering there is no creation or annihilation of particles.

Thus, if the standard model is indeed to emerge from quantum gravity, we require additional interactions beyond those considered in \cite{Bilson-Thompson:2006yc}.  This is not terrible as that paper studied a simplified form of dynamics  based on a limited class of local moves, connected naturally with graphs with trivalent nodes.  That these are not enough for quantum gravity is strongly suggested by the fact that in spin foam or path integral models of quantum spacetime geometry, the dynamics is associated with the natural local moves defined on tetra- and higher valent nodes.

As a result, recent investigations have focused on the issue of extending the dynamics studied
in \cite{Bilson-Thompson:2006yc,Hackett:2007dx} so that creation and annihilation of chiral excitations takes place while preserving precisely enough conservation laws to underlie the standard model.  One approach, suggested by \cite{fm-pc} has been to study the natural local moves of tetra-valent nodes.  This does lead to creation and annihilation of chiral excitations\cite{Wan2007,LeeWan2007,yidunjonathan}.  The approach taken here is complementary, we find topological invariants of a class of propagating braids whose conservation is consistent with the standard model interactions. This constitutes a second key result of this paper: any form of the local moves that does not preserve these topological invariants cannot yield the standard model interactions within the present scheme.  We do not, however, study in detail local moves of graphs which preserve these invariants. This is a task reserved for future work.

\begin{figure}[!h]
  \begin{center}
  \subfloat[]{\label{trin1}\includegraphics[scale=0.2]{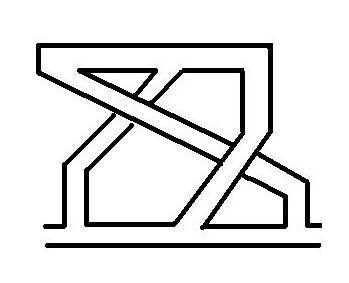}}
  \subfloat[]{\label{trin2}\includegraphics[scale=0.2]{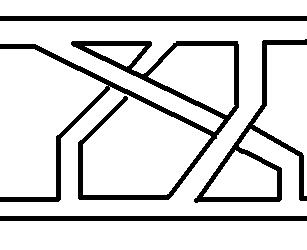}}
\end{center}
  \caption{Capped vs. Connected braids}
  \label{cpvscon}
\end{figure}

The methodology of this paper is as follows.  We begin by identifying a class of chiral braids which, as shown in \cite{Hackett:2007dx} do propagate on spin networks under trivalent local moves.  These are three stranded capped braids, illustrated in Figure (\ref{cpvscon}).  These were studied  in  \cite{LSM}, where it was shown that they can, in the orientable case, be classified in terms of a topological invariant consisting of triplets of integers $(a,b,c)$.   This classification is reviewed in section 2.

We then study, in section 3, the discrete symmetries of these excitations, resulting in the identification of analogues of $P,C$ and $T$,

In section 4 we give the main results of the paper.  We propose that there are  additional interactions which preserve the topological invariants $(a,b,c)$ and examine the consequences of this and the behavior of the excitations under
$C,P$ and $T$.  This leads to the identification of certain excitations with the weak vector bosons and the neutrinos.  If we also identify permutations on $(a,b,c)$ with permutations of the three colors we are led to the identifications of the quarks and leptons.  In fact, there is a discrete series of excitations which satisfy the conditions of the neutrinos, this leads to the structure of the generations.

The paper concludes with some comments about open issues in section 5.

Finally, we note that as in previous work in this series, the application to loop quantum gravity requires framed spin networks, which it has been
argued\cite{qdef} are necessary when the cosmological contant is non-zero.  There are no labels and no explicit evolution amplitudes in this paper, but the main results hold for any labeling of the graphs used in loop quantum gravity and any spin foam dynamics, so long as the conservation of $(a,b,c)$ is respected.

\section{Braided Ribbon Networks and twisting invariants}

Braided ribbon networks are a generalization from spin networks that introduce two complications: firstly that the edges are given a width and secondly that non-isotopic embeddings of the same spin network correspond to distinct basis states in a quantum space.  Restricting ourselves to trivalent spin networks gives us a two-surface composed of the union of trinions  (two-surfaces of the form shown in fig.\ref{trin1}) with the allowance of twists and braidings (fig. \ref{trin2}), embedded in a 3-manifold.

\begin{figure}[!h]
  \begin{center}
  \subfloat[]{\label{trin1}\includegraphics[scale=0.2]{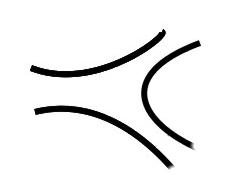}}
  \subfloat[]{\label{trin2}\includegraphics[scale=0.2]{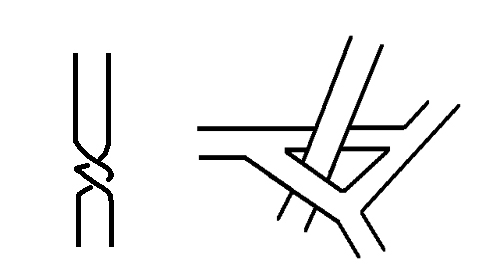}}
\end{center}
  \caption{Trinions, twists and braiding}
  \label{trinintro}
\end{figure}

\begin{figure}[!h]
  \begin{center}
    \includegraphics[scale=0.1]{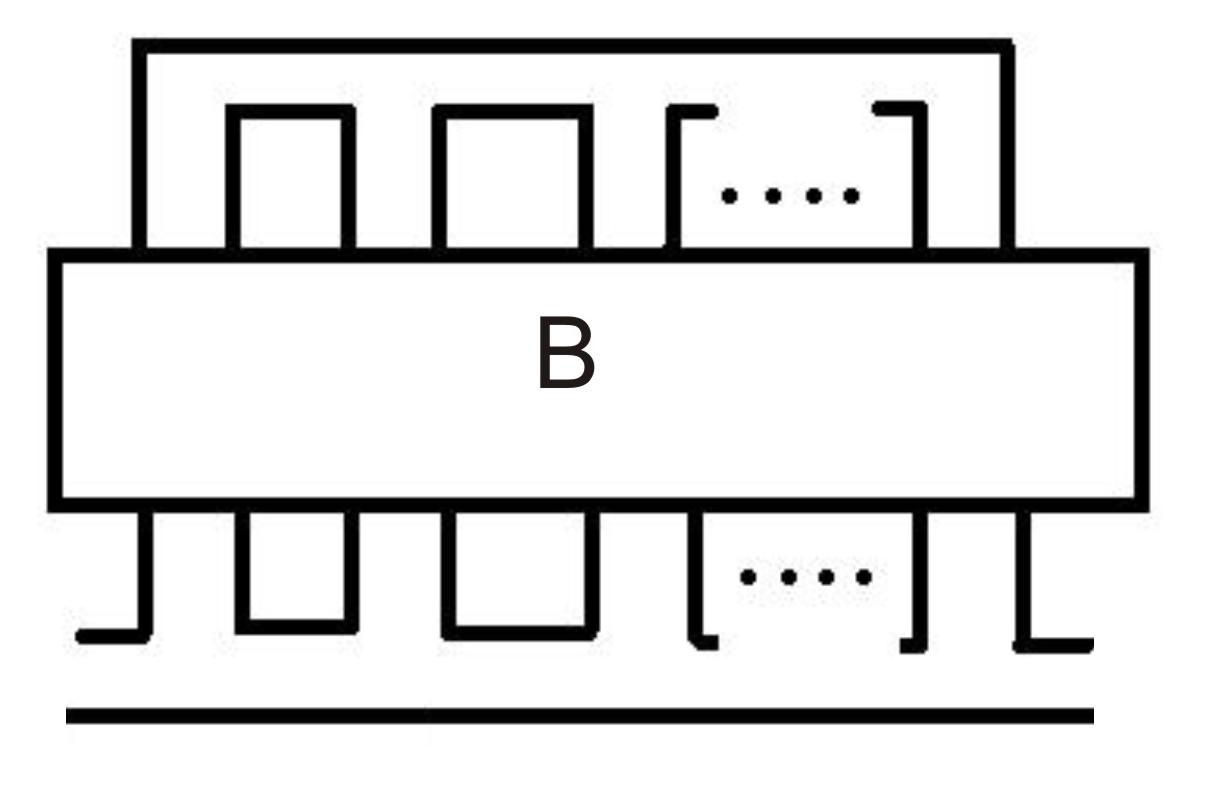}
  \end{center}
\caption{A capped n-braid} \label{nbraid}
\end{figure}

\begin{figure}[!h]
  \begin{center}
    \includegraphics[scale=0.05]{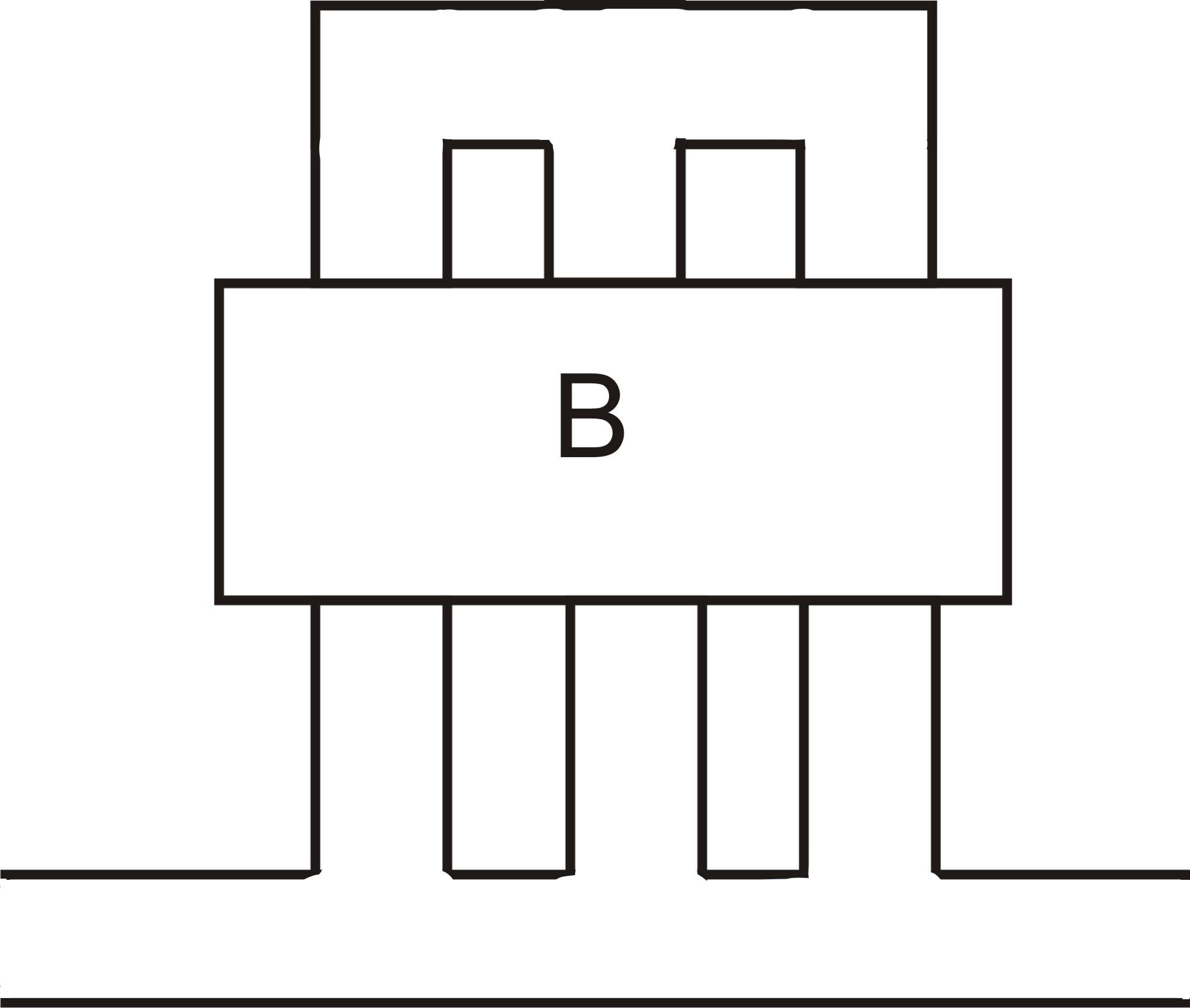}
  \end{center}
\caption{A capped 3-braid} \label{3braid}
\end{figure}

In \cite{Hackett:2007dx} it was proven that certain structures within braided ribbon networks were invariant under the standard evolution algebra $\mathcal{A}_{evol}$ of 3-valent spin networks, and certain classes of these were defined as isolatable sub-structures.  A subset of isolatable sub-structures are capped braids of $n$-strands (fig. \ref{nbraid}) 
where the B is an element of the braid group of $n$ strands $B_n$.  The simplest of these objects - and the one that is most natural in the setting of 3-valent networks - are capped $3$-braids (fig. \ref{3braid}).

\begin{figure}[!h]
  \begin{center}
  \subfloat[]{\label{nonor}\includegraphics[scale=0.2]{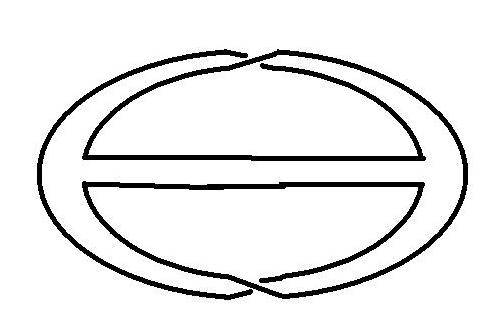}}
  \subfloat[]{\label{or}\includegraphics[scale=0.2]{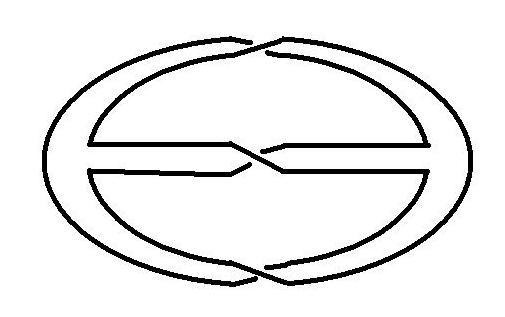}}
\end{center}
  \caption{Non-orientable and orientable Braided Ribbon Networks}
  \label{orient}
\end{figure}

We need to introduce one more concept before we proceed, that of orientability.  Braided ribbon networks include two types of surfaces: those where the surface has a distinct front and back (fig. \ref{or}), and those where it does not (fig. \ref{nonor}).  The first type of surface is called orientable and the second non-orientable.

\begin{figure}[!h]
  \begin{center}
  \subfloat[]{\label{braided}\includegraphics[scale=0.4]{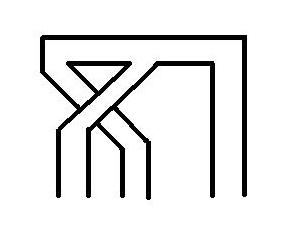}}
  \subfloat[]{\label{twisted}\includegraphics[scale=0.4]{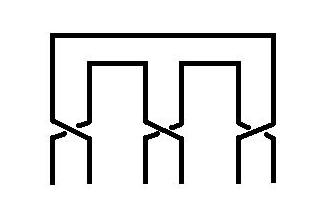}}
\end{center}
  \caption{Equivalence move by which braids are equivalent to twists}
  \label{braidntwist}
\end{figure}

As shown in \cite{LSM} each capped braid is equivalent to one with trivial 
braiding but various twists. This is illustrated in figure \ref{braidntwist}. 
Those three twists (ordered clockwise) are an invariant (a,b,c) by which we 
label states.  This invariant is generated by the fact that fig.\ref{braided} 
can be continuously deformed into fig.\ref{twisted} making the two surfaces 
equivalent. This same identification can be made for the other generators of 
the group $B_3$ and thus we can iteratively remove all braiding from a capped 
three braid. The twists on each of the three strands when all braiding has 
been removed define an invariant triplet of half-integers, $(a,\,b,\,c)$. 
Thus any capped three braid is deformable into any other capped three braid 
that shares the same invariant triplet. Additionally if we restrict our 
attention to orientable three braids we find that our vectors will not mix 
integers and half-integers.

\begin{figure}[!h]
  \begin{center}
    \includegraphics[scale=0.4]{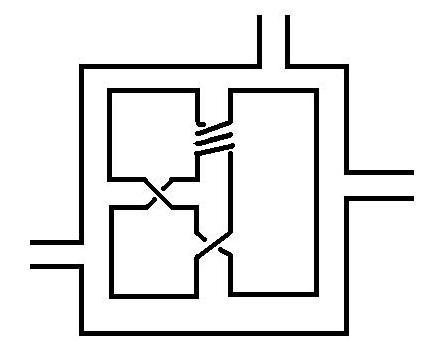}
  \end{center}
\caption{Invariant special case} \label{except}
\end{figure}

There is a single exception to the uniqueness of this invariant: situations 
where the ordering of the ribbons is not unique.  This exception is 
demonstrated by a situation where we can use the evolution algebra to isolate 
the braid in multiple ways each of which give a distinct ordering to the 
invariant (fig.\ref{except}), equivalent to making different choices for the 
left-most strand in a triplet.  This exception will not be further considered 
in this paper.

Capped $3$-braids cause further interest in their prior use in \cite{Bilson-Thompson:2006yc} to create a scheme introducing particle identifications into a quantum gravity setting.  The result in \cite{LSM} however makes us re-evaluate such a scheme in the context of capped $3$-braids as several of the initial attempts to extend the results in \cite{Bilson-Thompson:2006yc} treated different deformations of the same braid as candidates for distinct particles.

\section{Discrete symmetries of capped three braids and $C,P$ and $T$.}

In studying the invariants of capped three braids it is useful to determine the group of symmetries which govern them.  By considering the most general group we are initially inclined to consider $S_3$.  However, if we consider the fact that left-handed and right-handed twists are differentiated only by a minus sign within the invariant, we can also consider the two element group as an additional symmetry, giving $G = S_3 \times Z_2$.


We will write the group as follows:
\begin{quote}
Let $a \in S_3$ and let $+1$ and $-1$ be the two elements of $Z_2$ (with $-1$ being the element of order 2).  Then $a_{+1}$ and $a_{-1}$ $\forall a \in S_3$ will be the elements of $S_3 \times Z_2$.
\end{quote}

The group would act on the invariants as follows:
\begin{eqnarray}
(12)_{-1} \lhd [a,b,c] = [-b,-a,-c] \\
(123)_{+1}\lhd  [a,b,c] = [c,a,b]
\end{eqnarray}

The task then becomes to determine whether there is a subgroup of $G$ which yields appropriate elements corresponding to $C$, $P$ and $T$.  The choice of these elements will give us a foundation upon which to build a physical interpretation of these capped three braids.  Considering the idea that magnitude of charge should be invariant under these operations, we find that charge must take on the form of:
\begin{equation}
\mbox{Charge} = \chi (a+b+c)
\end{equation}
where $\chi$ is some function.  To respect convention we are led to suggest that $\chi$ be related to a factor of $\frac{1}{3}$.  As charge is additive, and so is twist, we are led to expect that $\chi$ involves only $(a+b+c)$, and no other powers this sum. The simplest choice is then that charge is $\frac{1}{3}(a+b+c)$. $C$ must then be an element of $S_3$ combined with a factor of $-1$.  Charge conjugation, parity and time reversal each have order 2, and should, for the case of states of massive particles at rest satisfy,

\begin{equation}
CPT = \bf{1}
\end{equation}
Taking this into account, our only option is for $\left( C,P,T \right)$ to be the set $\left( a_{-1}, a_{1}, \textbf{1}_{-1} \right)$ for some element $a \in S_3$, with order two.

Considering the three options of $(12)$, $(13)$ and $(23)$, $(13)$ becomes the preferred option as it corresponds to a physical operation that we can perform upon a Braided Ribbon Network as a whole.  A $(13)$ permutation can be generated across the entire network by looking at the network from `the other side' - this will reverse the ordering of the invariant of all braids.  Similarly a $(13)_{-1}$ can be achieved through a reflection of the network in its entirety, and a $\textbf{1}_{-1}$ by performing the two operations in succession.

Armed with this, we are left with our complete set of options: $C$ can only be $(13)_{-1}$ or $\textbf{1}_{-1}$, with $P$ and $T$ taking the other roles, as listed below.


We shall choose for the rest of the article the following definitions:

\begin{eqnarray}
C = \textbf{1}_{-1}\\
P = (13)\\
T = (13)_{-1}
\end{eqnarray}

additionally the charge of a braid defined by the invariant $[a,b,c]$ will be $\frac{1}{3}(a+b+c)$.

\section{Classification of Particle States}

We are now ready to present the main results of this paper, which is the complete scheme for the identification of the standard model fermions and weak vector bosons with excitations of quantum geometry.

\subsection{Invariant Classes}

By considering the action of the operators $C$, $P$ and $T$ upon classes of braids we can divide the braids into four categories.  The simplest category is that of braids which are invariant under both $C$ and $P$ (and hence also invariant under $T$).  These braids would then be equivalent to charge-less particles that do not have left- or right-handed versions (either scalars, or integer spin objects with secondary spin quantum number of zero).  A second class is objects which are invariant under $T$.  These objects necessarily must have the form $(a,0,-a)$.  We can then identify the two parity states in this situation with being left- and right-handed versions of the same object.
\begin{table}[h]
\begin{center}
\begin{tabular}{|c|c|c|c|} \hline
Class & Invariant & Number of Corresponding Objects & Invariant under \\\hline
 I & $[0,0,0]$ & One & $C$,$P$,$T$\\
 II & $[a,0,-a]$ & Two & $T$ \\
 III & $[a,b,a]$ & Two & $P$ \\
 IV & $[a,b,c]$ & Four & None\\
 \hline
\end{tabular}
\end{center}
\caption{Classes of Braids}
\label{class}
\end{table}

Table \ref{class} demonstrates these categories.  We can see that the categories discussed above fall exactly into class I and II.  This leaves the other two categories to explain.  Class III braids are then in correspondence with particles that are invariant under parity - charged scalars and objects with secondary spin quantum number of zero (i.e. the $m_s = 0$ states of integer spin objects).   It is interesting to note that Class II braids respond to $C$ and $P$ in the same manner, implying that their parity transforms are also their charge conjugates.  Class IV objects are then sets of four braids that map to one another under $C$,$P$ and $T$ corresponding with fermions.

\subsection{Interaction Assumptions}

We will assume that there is some mechanism that allows interactions between braids in the three forms of:

\begin{eqnarray}
\left[a,b,c\right]+\left[d,e,f\right] = \left[a+d,b+e,c+f\right]\\
\left[a,b,c\right] = \left[d,e,f\right]+\left[g,h,i\right],\ \ \  d+g = a,\ \ \  e+h=b,\ \ \  f+i = c\\
\left[a,b,c\right]+\left[d,e,f\right] = \left[g,h,i\right]+ \left[j,k,l\right],\ \ \  a+d=g+j,\ \ \  b+e=h+k,\ \  \ c+f=i+l
\end{eqnarray}

\subsection{The Z}

From this, we make a choice for the Z-boson.  As the Z is a spin-1 particle without a charge conjugation pair, it falls into the category of class I \& II braids.  Adding in a further requirement that we desire our Z-boson to be representable by a braid which does not contain any twists, we find that our first successful candidate for it is the $[2,0,-2]$ state (the $m_s = 0$ state necessarily becomes the $[0,0,0]$ state).  The $W^+$ and $W^-$ states should then take on the form of the Z states with an extra integer twist placed on each strand, giving a set of eight objects (these are explicitly shown in table \ref{biglist}).

\subsection{The Neutrino}

We shall require that a neutrino be a particle, represented by a braid which 
does not contain any twists, with a neutral charge.  Additionally it should 
satisfy that its parity conjugate is equal to its interaction with one of the 
Z bosons (i.e. given the evidence that the neutrinos of at least two 
generations are massive, we expect that right-handed neutrinos exist, and 
will interact with left-handed Z bosons to produce left-handed neutrinos).  
We find a countable number of such particles satisfying:

\begin{equation}
\left[a,2-2a,a-2 \right] ,\ \ \ a \in 2Z
\end{equation}

The first such example is then $\left[2,-2,0]\right]$ which corresponds 
to a left-handed neutrino.  It follows that all such objects are class IV 
braids, and therefore come in sets of four objects.  This implies that 
there are both left-handed and right-handed neutrinos in all generations 
and that there are an infinite series of such neutrinos.

\subsection{The Rest of the Scheme}
\begin{table}[h]

\begin{center}

\begin{tabular}{|llll|llll|} \hline

\multicolumn{4}{|c|}{Particle} & \multicolumn{4}{|c|}{Invariant} \\ \hline

$Z_L\hspace{5mm}$&   $Z_R\hspace{5mm}$&   $Z_0\hspace{5mm}$&   &
$[2,0,-2]\hspace{5mm}$&  $[-2,0,2]\hspace{5mm}$&  $[0,0,0]\hspace{5mm}$& \\

$W^+_L$& $W^+_R$& $W^+_0$& & $[3,1,-1]$&  $[-1,1,3]$&  $[1,1,1]$& \\

$W^-_L$& $W^-_R$& $W^-_0$& & $[1,-1,-3]$& $[-3,-1,1]$& $[-1,-1,-1]$& \\

\hline

$\nu^{(e)}_L$& $\nu^{(e)}_R$& $\bar{\nu}^{(e)}_L$& $\bar{\nu}^{(e)}_R$&
  $[2,-2,0]$& $[0,-2,2]$& $[-2,2,0]$& $[0,2,-2]$ \\

$e^{-}_L$& $e^{-}_R$& $e^{+}_L$& $e^{+}_R$& $[1,-3,-1]$& $[-1,-3,1]$& $[-1,3,1]$& $[1,3,-1]$ \\

$d_{L,r}$& $d_{L,g}$& $d_{L,b}$& & $[1,-2,0]$& $[2,-3,0]$& $[2,-2,-1]$& \\

$d_{R,r}$& $d_{R,g}$& $d_{R,b}$& & $[-1,-2,2]$& $[0,-3,2]$& $[0,-2,1]$& \\

$\bar{d}_{L,r}$& $\bar{d}_{L,g}$& $\bar{d}_{L,b}$& & $[-1,2,0]$& $[-2,3,0]$& $[-2,2,1]$&  \\

$\bar{d}_{R,r}$& $\bar{d}_{R,g}$& $\bar{d}_{R,b}$& & $[1,2,-2]$& $[0,3,-2]$& $[0,2,-1]$& \\

$u_{L,r}$& $u_{L,g}$& $u_{L,b}$& & $[2,-1,1]$& $[3,-2,1]$& $[3,-1,0]$& \\

$u_{R,r}$& $u_{R,g}$& $u_{R,b}$& & $[0,-1,3]$& $[1,-2,3]$& $[1,-1,2]$& \\

$\bar{u}_{L,r}$& $\bar{u}_{L,g}$& $\bar{u}_{L,b}$& & $[-2,1,-1]$& $[-3,2,-1]$& $[-3,1,0]$&  \\

$\bar{u}_{R,r}$& $\bar{u}_{R,g}$& $\bar{u}_{R,b}$& & $[0,1,-3]$& $[-1,2,-3]$& $[-1,1,-2]$& \\

\hline

$\nu^{(\mu)}_L$& $\nu^{(\mu)}_R$& $\bar{\nu}^{(\mu)}_L$& $\bar{\nu}^{(\mu)}_R$& $[4,-6,2]$& $[2,-6,4]$& $[-4,6,-2]$& $[-2,6,-4]$\\

$\mu_L$& $\mu_R$& $\bar{\mu}_L$& $\bar{\mu}_R$& $[3,-7,1]$& $[1,-7,3]$& $[-3,7,-1]$& $[-1,7,-3]$\\

$c_{L,r}$& $c_{L,g}$& $c_{L,b}$& & $[4,-5,3]$& $[5,-6,3]$& $[5,-5,2]$& \\

$c_{R,r}$& $c_{R,g}$& $c_{R,b}$& & $[2,-5,5]$& $[3,-6,5]$& $[3,-5,4]$& \\

$\bar{c}_{L,r}$& $\bar{c}_{L,g}$& $\bar{c}_{L,b}$& & $[-4,5,-3]$& $[-5,6,-3]$& $[-5,5,-2]$& \\

$\bar{c}_{R,r}$& $\bar{c}_{R,g}$& $\bar{c}_{R,b}$& & $[-2,5,-5]$& $[-3,6,-5]$& $[-3,5,-4]$& \\

$s_{L,r}$& $s_{L,g}$& $s_{L,b}$& & $[5,-6,2]$& $[4,-7,2]$& $[4,-6,1]$& \\

$s_{R,r}$& $s_{R,g}$& $s_{R,b}$& & $[1,-6,4]$& $[2,-7,4]$& $[2,-6,3]$& \\

$\bar{s}_{L,r}$& $\bar{s}_{L,g}$& $\bar{s}_{L,b}$& & $[-3,6,-2]$& $[-4,7,-2]$& $[-4,6,-1]$& \\

$\bar{s}_{R,r}$& $\bar{s}_{R,g}$& $\bar{s}_{R,b}$& & $[-1,6,-4]$& $[-2,7,-4]$& $[-2,6,-3]$&  \\ \hline

\end{tabular}

\caption{Classes of Braids}

\label{biglist}

\end{center}
\end{table}

Further particles are found by adding twists to the individual strands of the neutrinos (without double twisting a single strand), the colour of a quark being determined by the `odd strand out' (the untwisted strand when two strands are twisted, or the only twisted strand when a single one is).  This twisting gives them their expected charges, and the colouring gives the expected results for interactions with the $W$'s. Table \ref{biglist} shows the first two generations of particles, along with the bosons.

This pattern then continues for infinitely many higher generations, each made from successive neutrino states.

\section{Conclusion}

We have presented an embedding of the fermion and weak vector boson states of the standard model in a class of loop quantum gravity models.  These are models in which the states are based on embeddings of framed trivalent spin networks, with possibly arbitrary labellings, whose dynamics is given by the standard dual Pachner trivalent moves, plus additional moves consistent with the conservation of the topological invariants $(a,b,c)$.

There are a number of interrelated questions that remain open before the promise of this development can be fully understood. 

\begin{itemize}

\item{} We have not attempted to identify states associated with the unbroken gauge invariances of the standard model, namely $SU(3)\times U(1)$.   We believe these gauge interactions may be emergent from the dynamics of the states described here; this is an old idea, which would need to be developed in the present context.  

\item{} The low energy dynamics of these excitations should be computable by seeing them as excitations of semiclassical states of loop quantum gravity.  In particular, the dynamical breaking of the chiral symmetry leading to a spectrum of fermion masses needs to be studied.   Of particular interest will be the plausibility of there being additional generations whose masses, particularly the neutrinos, evade present limits.  Of course, were evidence for additional generations discovered at the $LHC$ that would support the theory described here. 

\item{}There are apparently exotic excitations which would be predictions for new states beyond the standard model, which need to be explored.

\item{}It remains the case that the states we have identified as fermions have not been shown to be either spinorial or fermionic.  As mentioned in 
\cite{Bilson-Thompson:2006yc}, this may be understood by as a kind of inverse quantum Hall effect, but this has not so far been further developed.

\end{itemize}

While there is much remaining to be done, the possibility that the observed elementary particles are excitations of quantum geometry remains worthy of further development.  



\section{Acknowledgements}

We would like to thank Fotini Markopoulou and Yidun Wan for discussions and comments during the course of this work.  
Sundance Bilson-Thompson wishes to thank the Special Research Centre for the Subatomic Structure of Matter at the University of Adelaide for their kind hospitality during part of the preparation of this manuscript. Research at Perimeter Institute for Theoretical Physics is supported in part by the Government of Canada through NSERC and by the Province of Ontario through MRI.


\begin{thebibliography}{99}

\bibitem{Pati:1974yy}
  J.~C.~Pati and A.~Salam,
  Phys.\ Rev.\  D {\bf 10}, 275 (1974)
  [Erratum-ibid.\  D {\bf 11}, 703 (1975)].

\bibitem{Terazawa:1980hh}
  H.~Terazawa and K.~Akama,
  Phys.\ Lett.\  B {\bf 96}, 276 (1980).

\bibitem{Harari:1979gi}
  H.~Harari,
  Phys.\ Lett.\  B {\bf 86}, 83 (1979).

\bibitem{Shupe:1979fv}
  M.~A.~Shupe,
  Phys.\ Lett.\  B {\bf 86}, 87 (1979).

\bibitem{Harari:1980ez}
  H.~Harari and N.~Seiberg,
  Phys.\ Lett.\  B {\bf 98}, 269 (1981).

\bibitem{Bilson-Thompson:2005bz}
  S.~O.~Bilson-Thompson,
``A topological model of composite preons,''
  arXiv:hep-ph/0503213.

\bibitem{Bilson-Thompson:2006yc}
  S.~O.~Bilson-Thompson, F.~Markopoulou and L.~Smolin,
  ``Quantum gravity and the standard model,''
  Class. Quant. Grav. {\bf 24} (2007) 3975-3993, 
  arXiv:hep-th/0603022.

\bibitem{Hackett:2007dx}
  J.~Hackett,
  ``Locality and translations in braided ribbon networks,''
  arXiv:hep-th/0702198.

\bibitem{LSM} S.~O.~Bilson-Thompson, J.~Hackett, L.~Kauffman and L.~Smolin In Preperation.

\bibitem{fot-dav}D. W. Kribs and F. Markopoulou, arXiv:gr-qc/0510052;
 F. Markopoulou, {\it Towards Gravity from the Quantum},   arXiv:hep-th/0604120.

\bibitem{nfs}P.Zanardi and M.Rasetti, Phys.Rev.Lett.79 3306 (1997);
D. A. Lidar, I. L. Chuang and K. B. Whaley, Phys. Rev. Lett. 81, 2594 (1998)
[arXiv:quant-ph/9807004];
E.Knill, R.Laflamme and L.Viola, Phys.Rev.Lett. 84 2525 (2000);
J. Kempe, D. Bacon, D. A. Lidar and K. B. Whaley, Phys. Rev. A 63, 042307 (2001).

\bibitem{fm-pc}F. Markopoulou, personal communication.

\bibitem{Wan2007}Y. Wan, ``On Braid Excitations in Quantum Gravity'', arXiv:0710.1312.

\bibitem{LeeWan2007}L. Smolin, Y. Wan,
``Propagation and Interaction of chiral states in quantum
gravity'', arXiv:0710.1548, accepted for publish in Nucl. Phys. B.

\bibitem{yidunjonathan}J. Hackett and Y. Wan,  {\it Conserved Quantities for Interacting Four Valent Braids in Quantum Gravity}, arXiv:0803.3203. 



\bibitem{qdef}S. Major and L. Smolin, Quantum deformation of quantum gravity, Nucl. Phys. B473,
267(1996), gr-qc/9512020; R. Borissov, S. Major and L. Smolin, The geometry of quantum spin networks, Class. and Quant. Grav.12, 3183(1996), gr-qc/9512043; L. Smolin, 
Quantum gravity with a positive cosmological constant, hep-th/0209079.

\end{thebibliography}
\end{document}